# Constructing Berry-Maxwell equations with Lorentz invariance and Gauss's law of Weyl monopoles in 4D energy-momentum space

Yiming Pan[1], Ruoyu Yin[2]

1. School of Physical Science and Technology and Center for Transformative Science, ShanghaiTech University, Shanghai 200031, China

2. Department of Physics, Institute of Nanotechnology and Advanced Materials, Bar-Ilan University, Ramat-Gan 52900, Israel

Email: yiming.pan@shanghaitech.edu.cn

**Abstract**

We present the construction of a reciprocal electromagnetic field by extending the Berry curvatures into four-dimensional (4D) energy-momentum space. The resulting governing equations, termed Berry-Maxwell equations, are derived, by incorporating Lorentz invariance to constrain the parameter space of energy-momentum. Notably, these Berry-Maxwell equations exhibit dual and self-dual structures compared to the Maxwell equations. The very existence of Berry-Maxwell equations is independent of the geometrical phase of matter waves, implying that they cannot be directly derived from the time-dependent Schrödinger equation. Indeed, we find that the physical reality of this reciprocal electromagnetic field is rooted in the fundamental principles of special relativity and Gauss's law of Weyl monopoles. To validate our theory experimentally, we outline three effects for verification: (i) Lorentz boost of a Weyl monopole, (ii) reciprocal Thouless pumping, and (iii) plane-wave solutions of Berry-Maxwell's equations.

Dualities play a prominent role in modern physics, manifesting in various theories and phenomena. Notable examples include the wave-particle duality and Born reciprocity in quantum mechanics [1,2], the duality between electric and magnetic fields in Maxwell equations and its generalizations [3–7], the Sine-Gordon/Thirring duality [8–11], Krammers-Wannier duality in Ising models [12,13], gauge and gravity duality in string theories [14–16]. Typically, duality arises when seemingly disparate aspects of nature are connected through the reconstruction or rearrangement of quantities in a theory or between two different theories. These dualities often offer alternative approaches for studying nonperturbative behaviors at strong coupling regime [16]. In the context of the debate on reductionism and emergence, understanding dual physical quantities or theories provides a perspective of "democracy among particles" [10]. It is important to note that while duality is a compelling concept, it lacks a well-defined physical interpretation comparable to the symmetries that govern the dynamics of a system. Instead, it appears to be more of a metaphysical or philosophical similarity describing exact (or up to a certain limit) mathematical structures shared by two or more apparently different theories [17,18].

In condense matter physics, the quantum Hall effects [19–21] and topological insulators [22,23] have emerged as compelling demonstrations of Berry curvatures in the generalized parameter space of physical systems. Initially met with skepticism in quantum mechanics, the closely associated geometric phases have since undergone significant development in understanding topological matters. Notably, striking analogies arise between the Berry connection, Berry curvature, Berry phase and Chern number in momentum space, and the vector potential, static magnetic field, Aharonov-Bohm phase, and quantized magnetic flux in real space, respectively [24,25]. This apparent analogy between Berry curvature ($\Omega$) and magnetic field ($B$) raises intriguing questions: (i) Is it possible to construct an electric-like Berry curvature ($\Upsilon$) in analogous to the real-space electric field ($E$)? (ii) If such an analogy can be established, can we then connect the electric- and magnetic-like Berry curvatures to formulate a set of coupled equations, akin to the celebrated Maxwell equations? Moreover, if these equations do exist, would they give rise to the following dualities:

$$\begin{aligned}\Upsilon \to \Omega, \Omega \to -\Upsilon &: \text{Electric/magnetic duality of Berry curvatures,} \\ \Upsilon \leftrightarrow E, \Omega \leftrightarrow B &: \text{Berry/Maxwell duality?}\end{aligned} \qquad (1)$$

The first expression illustrates the self-duality between Berry curvatures that is similar to the conventional duality between electric and magnetic fields (E → B, B → −E), while the second expression highlights the duality between our constructed Berry-Maxwell equations and the known Maxwell equations. This implies that the structure of Berry-Maxwell equations remains invariant under self-dual and dual transformations (1). Exploiting these dual structures in conjunction with Maxwell equations, we can conjecture the formulation of Berry-Maxwell equations of Berry curvatures in 4D energy-momentum space.

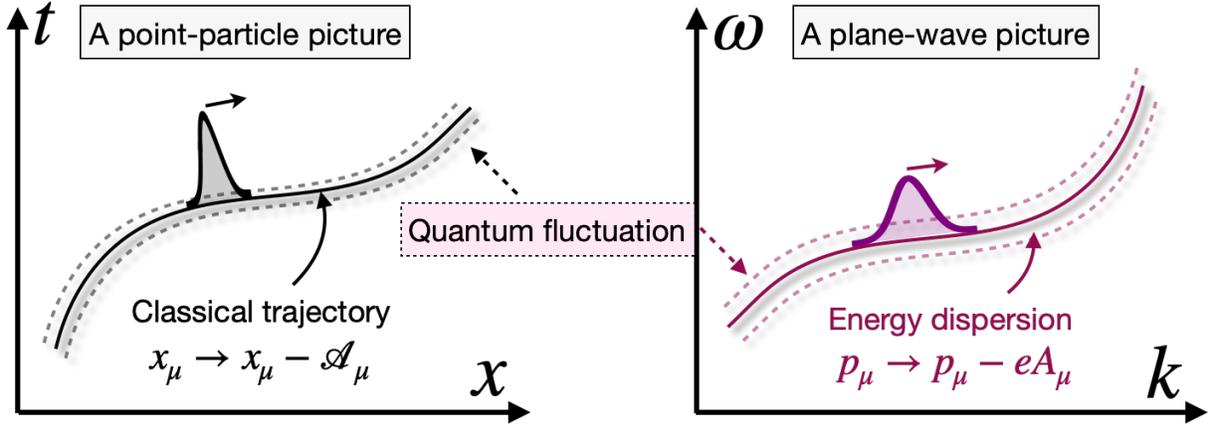

Fig. 1: Born reciprocity between classical trajectory of a particle and energy dispersion of a wave. The quantum fluctuation of this reciprocity in phase space are bounded by the Heisenberg's uncertainty principle. When a wavepacket moves in phase space in the presence of the electromagnetic field or the reciprocal electromagnetic field, it can be effectively described through the substitutions depicted as (left panel) $x_\mu \rightarrow x_\mu - \mathcal{A}_\mu$ or (right panel) $p_\mu \rightarrow p_\mu - eA_\mu$, respectively, in which $\mathcal{A}_\mu$ is the Berry four-connection and $A_\mu$ is the electromagnetic four-potential. Notice that the classical trajectory $x = x(t)$ leads to a constraint in spacetime for a particle's dynamics, and the energy-momentum dispersion $\omega = \omega(k)$ leads to a constraint in energy-momentum space for a wave equation. In our construction, these constraints are violated by quantum fluctuation, so the fields $\mathcal{A}_\mu$ and $A_\mu$ are independently defined.

Indeed, a set of Berry-Maxwell equations exists, as demonstrated in (2). The explicit construction is as follows: First, we extend the Berry connection and Berry curvature into 4D energy-momentum space, resulting in the formulation of the Berry four-connection. This extension gives rise to an additional electric-like Berry curvature, termed the "reciprocal electric field". Second, utilizing the Born reciprocity [2] between the Berry four-connection and electromagnetic four-potential, we can construct the Berry-Maxwell equations governing the possible dynamics of the reciprocal electromagnetic field. To achieve this, we introduce Lorentz invariance for the 4D energy-momentum space and the Gauss's law for a Weyl point acting as a magnetic monopole [26,27]. Starting with the Weyl monopole, we can explicitly derive the Berry-Maxwell equations:

$$\begin{aligned} \nabla_k \cdot \Upsilon &= 0 \\ \nabla_k \cdot \Omega &= \rho_W \\ \nabla_k \times \Upsilon &= -\frac{\partial \Omega}{\partial \omega} - j_W \\ \nabla_k \times \Omega &= \frac{\partial \Upsilon}{\partial \omega} \end{aligned} \qquad (2)$$

with the definitions $\Omega = \nabla_k \times \mathcal{A}(k, \omega), \Upsilon = -\frac{\partial}{\partial \omega}\mathcal{A}(k, \omega) - \nabla_k \chi(k, \omega)$, representing the magnetic- and electric-like Berry curvatures, respectively. The Berry four-connection $(\chi, \mathcal{A})$ is a direct extension of the Berry connection $(\mathcal{A})$ in 3D momentum space. The magnetic-like density $\rho_W$ corresponds to the density of the Weyl monopoles, while the term $j_W$ represents the magnetic-like current flow of Weyl monopoles. This current flow follows an analogy of a continuity equation (Eq. 12), i.e., $\frac{\partial}{\partial \omega}\rho_W + \nabla_k \cdot j_W = 0$. The explicit expressions will be constructed in the following sections. In contrast, the conventional Maxwell's equations in the energy-momentum space take the form

$$\begin{cases} i\boldsymbol{k} \cdot E = \frac{\rho}{\epsilon_0} \\ i\boldsymbol{k} \cdot B = 0 \\ i\boldsymbol{k} \times E = i\omega B \\ i\boldsymbol{k} \times B = -\frac{i\omega}{c^2}E - \frac{i\omega}{\epsilon_0 c^2}j \end{cases} \qquad (3)$$

in terms of plane-wave solutions with $\rho$ denoting the electric charge and $j$ representing the electric current. The dual and self-dual structures between the Berry-Maxwell equations (2) and the Maxwell equations (3) in absence of sources or currents can be readily verified. However, it is crucial to emphasize that these two fields exist independently. In Fig. 1, a quantum matter wave can be described either as a point-particle in spacetime in the classical limit or as a plane-wave in energy-momentum in the quantum limit. When the quantum system interacts with the Maxwell and Berry-Maxwell fields, the corresponding interaction forms are determined through the generalized minimal couplings: $p_\mu \to p_\mu - eA_\mu, x_\mu \to x_\mu - A_\mu$. Notably, quantum fluctuations between spacetime and energy-momentum space are subjected to Heisenberg's uncertainty principle. Furthermore, we present the construction of generalized Lorentz equations in phase space in Appendix 5. These equations can be regarded as a specific manifestation of an eight-dimensional Berry curvature construction, grounded in the semiclassical wave packet's motion in phase space, as proposed by Q. Niu et al. [28,29].

Finally, we investigate three nontrivial effects and consequences stemming from our construction of Berry-Maxwell equations. (i) We note the absence of electric-like source in our formulation (Eq. 2), precluding the direct derivation of the nontrivial electric-like Berry curvature. Nonetheless, we demonstrate that the electric-like Berry curvature can be induced by the Lorentz boost of a Weyl monopole [30–32] in a moving frame. (ii) We perform a comparative analysis of various Chern numbers and their physical realizations, revealing that the Chern number in energy-momentum space exhibits behavior akin to a Thouless pumping [33,34]. We term this intriguing phenomenon "reciprocal Thouless pumping", which remains unexplored in experimental observations. (iii) We derive plane-wave solutions for the Berry-Maxwell equations even in the absence of Weyl monopoles or currents. Remarkably, we ascertain that the plane-wave solution in our construction represents an exotic point-like event in spacetime. In light of these findings, we envision that our construction will unveil further dualities within quantum mechanics, and we anticipate that the electric-like Berry curvature and Berry-Maxwell equations will lead to many nontrivial physical consequences awaiting for experimental verifications.

**Generalization of Berry curvature in 4D energy-momentum space.** To be specific, we construct the Berry four-connection associated with the Floquet-Bloch eigenstates of a spacetime crystal quantum system. Considering a spacetime crystal with a spatiotemporally periodic Hamiltonian, the time-dependent Schrodinger equation (TDSE) is

$$i\hbar \frac{\partial}{\partial t} \psi(x,t) = \left( \frac{(-i\hbar \nabla)^2}{2m} + V(x,t) \right) \psi(x,t) \quad (4)$$

where the periodic potential $V(x,t) = V(x+a, t+T)$ varies in space and time, with lattice constant $a$ and period T. Applying the Floquet-Bloch theorem,

$$\psi_n(x,t) = e^{ik \cdot x - i\omega t} u_n(x,t) \tag{5}$$

where k is the Bloch quasi-momentum, $\omega$ is Floquet quasi-energy, and n is the Floquet-Bloch band index. For simplicity, we suppress the band index notation in the Berry connection and curvature below. The Floquet-Bloch state $u_n(x,t) = u_n(x+a, t+T)$ satisfies $H_F |u_n(x,t)\rangle = \omega |u_n(x,t)\rangle$ with the Floquet Hamiltonian defined as $H_F = H - i\hbar \partial_t$. According to our general formulation (A4), the Berry connections of the n$^{th}$ Floquet-Bloch band in energy-momentum space are given by

$$\begin{aligned} \chi_n(k,\omega) &= +i\langle u_n(k,\omega,x,t)|\partial_\omega|u_n(k,\omega,x,t)\rangle \\ \mathcal{A}_n(k,\omega) &= -i\langle u_n(k,\omega,x,t)|\partial_k|u_n(k,\omega,x,t)\rangle \end{aligned} \tag{6}$$

The corresponding Berry curvatures are defined as follows:

$$\begin{aligned} \Upsilon &= -\frac{\partial}{\partial \omega}\mathcal{A}(k,\omega) - \nabla_k \chi(k,\omega), \\ \Omega &= \nabla_k \times \mathcal{A}(k,\omega). \end{aligned} \tag{7}$$

Here, $\Upsilon$ represents the electric-like Berry curvature and $\Omega$ denotes a magnetic-like Berry curvature. The electric-like Berry curvature has been discussed as an artificial electric field in an interacting Fermi liquid [35]. Analogous to the electromagnetic field tensor, we express the Berry curvature as a "reciprocal electromagnetic field tensor", given by

$$\Omega_{\mu\nu} = \partial_\mu \mathcal{A}_\nu - \partial_\nu \mathcal{A}_\mu = \begin{pmatrix} 0 & +\Upsilon_x & +\Upsilon_y & +\Upsilon_z \\ -\Upsilon_x & 0 & -\Omega_z & +\Omega_y \\ -\Upsilon_y & +\Omega_z & 0 & -\Omega_x \\ -\Upsilon_z & -\Omega_y & +\Omega_x & 0 \end{pmatrix}. \tag{8}$$

The electric Berry connection (6) is related to the magnetic Berry connections through the energy-momentum dispersion constraint $\omega = E(k)$. To relax this constraint, A. Li et al. [29], introduced a "proper time" $\tau$ in the geodynamic equation: $\mathcal{L}(-i\partial_x, i\partial_t; x,t)\Psi(x,t) = i\hbar \frac{\partial}{\partial \tau} \Psi(x,t)$, where $\mathcal{L}$ is the Floquet Hamiltonian $H_F$. By substituting $\Psi(x,t) = e^{-\frac{i\lambda_0 \tau}{\hbar}} \psi(x,t)$, we obtain the eigenproblem: $\mathcal{L}(-i\partial_x, i\partial_t; x,t)\psi = \lambda_0 \psi$. Applying the Floquet-Bloch theorem again yields a dispersion function $\lambda_0 = \lambda_0(k,\omega)$, a scale function of energy and momentum. The corresponding off-shell Floquet-Bloch states $u_{\lambda_0}(x,t)$ are given by $\mathcal{L}(-i\partial_x + k, i\partial_t - \omega; x, t)u_{\lambda_0}(x,t) = \lambda_0 u_{\lambda_0}(x,t)$. The original wave equation corresponds to the zero "energy" solution: $\lambda_0 = 0$. Indeed, the off-shell state $u_{\lambda_0}(x,t)$ is analytical similar to the on-shell state $u_n(x,t)$ but with the frequency $\omega$ shifted by $\lambda_0$. Further details can be found in Appendix 5 and [29].

On the other hand, for the time-independent system, we can further express the magnetic Berry curvature $\Omega_{\mu\nu}$ in terms of $\Omega_{\mu\nu} = i\langle \partial_{R_\mu}\psi_n | \partial_{R_\nu}\psi_n \rangle - i\langle \partial_{R_\nu}\psi_n | \partial_{R_\mu}\psi_n \rangle$, as insert that the completeness $\sum_{n'} |\psi_{n'}\rangle\langle\psi_{n'}| = I$, we then obtain the spectral representation:

$$\Omega_{\mu\nu} = i \sum_{n'} \langle \partial_{R_\mu}\psi_n | \psi_{n'} \rangle \langle \psi_{n'} | \partial_{R_\nu}\psi_n \rangle - (\mu \leftrightarrow \nu)$$
$$= i \sum_{n' \neq n} \frac{\langle n | \partial_{R_\mu} H | n' \rangle \langle n' | \partial_{R_\nu} H | n \rangle}{(E_n - E_{n'})^2} - (\mu \leftrightarrow \nu). \quad (9)$$

Notice that the term $n' = n$ is cancelled out due to the substitution $(\mu \leftrightarrow \nu)$, where $n, n'$ represent an abbreviated notation for Bloch eigenstates $\psi_n, \psi_{n'}$. It is important to note that the Berry curvature can be properly defined for non-adiabatic Floquet steady states when the parameter space of Bloch basis is extended to the Floquet-Bloch basis. Constructing the generalized Berry phase (aka, Aharonov-Anandan phase [36]) does not necessarily require adiabatic condition, instead a cyclic evolution is sufficient to define the geometric phase. The Berry curvature diverges at the degenerate point where $E_n = E_{n'}$. To explore the power law of this divergence, we linearize the denominator at the degenerate point when $R = R_c$: $E_n - E_{n'} = \left(\frac{\partial E_n}{\partial R} - \frac{\partial E_{n'}}{\partial R}\right)(R - R_c)$. Given that $\left(\frac{\partial E_n}{\partial R} - \frac{\partial E_{n'}}{\partial R}\right) \neq 0$ and the numerator is nonzero, we can expect the relationship: $\Omega \sim \frac{1}{(R-R_c)^2}$. Consequently, we derive a nontrivial Gauss' law for this divergence of Berry curvature: $\nabla_R \cdot \Omega \sim \delta^3(k - k_c)$.

It is noteworthy that at the degenerate point where $E_n = E_{n'}$, the adiabatic approximation fails, resulting in a singularly ill-defined magnetic-like Berry connection. However, the Berry curvature still remains comprehensible by treating the divergence as a magnetic monopole in momentum space. For instance, consider the example of a 3D Weyl point, characterized by a Hamiltonian $H_{eff} = v_F k \cdot \sigma$, representing a level crossing of two nondegenerate energy bands (see Ref. [37]). Explicitly, the magnetic-like Berry curvature is given by

$$\Omega = \nabla_k \times \mathcal{A} = -\frac{\hat{e}_k}{2k^2} \tag{10}$$

Considering a monopole located at the origin, we define the scalar potential $\tilde{\mathcal{V}}(k) = -\frac{1}{2k}$, and subsequently take the gradient of $\tilde{\mathcal{V}}(k)$ in the momentum space, namely, $\Omega = -\nabla_k \tilde{\mathcal{V}}(k) = -\frac{\partial}{\partial k}\left(-\frac{1}{2k}\right) = -\frac{\hat{e}_k}{2k^2}$. Notably, $\Omega$ can be expressed as the negative gradient of a scalar potential, leading to $\nabla_k \times \Omega = 0$, since the curl of a gradient always equals zero. At the moment, we can derive the monopole density, given by $\rho_W = \nabla_k \cdot \Omega = 2\pi\, \delta^3(k)$.

Note that incorporating higher-order terms in the expansion of $E_n = \alpha_n p + \beta_n p^2 + \gamma_n p^3 + \cdots$ can unveil the contribution of high-order energy momentum dispersion to the Berry curvature. This expansion reveals that around the critical point $R_c$, the higher-order terms are negligible. As one moves away from $R_c$, the divergence of Berry curvature manifests an inhomogeneous magnetic polarization effect. Such a polarized monopole density would be inserted into the Berry-Maxwell equations as an inhomogeneous reciprocal source, which offers us various magnetic induction to explore the reciprocal electromagnetic field.

**Proof of Berry-Maxwell equations.** Inspired from the derivation of Maxwell equations through the marriage of special relativity and Gauss's law of electric charge (see Appendix 4), we introduce the Lorentz invariance into the Berry curvature. We consider a moving frame S' with speed $v$ in x-direction relative to the frame S at rest, where the coordinate systems are related by the Lorentz boost: $x' = \gamma(x - \beta ct), ct' = \gamma(ct - \beta x)$, while $y' = y, z' = z$ remain unchanged. Similarly, the energy-momentum space undergoes the same Lorentz boost. That is to say, we can imprint the constraint of Lorentz invariance from spacetime observables into the quantities in energy-momentum space. The corresponding Lorentz transformation is given by

$$\begin{pmatrix} \omega'/c \\ k'_x \\ k'_y \\ k'_z \end{pmatrix} = \begin{pmatrix} \gamma & -\beta\gamma & 0 & 0 \\ -\beta\gamma & \gamma & 0 & 0 \\ 0 & 0 & 1 & 0 \\ 0 & 0 & 0 & 1 \end{pmatrix} \begin{pmatrix} \omega/c \\ k_x \\ k_y \\ k_z \end{pmatrix} = \begin{pmatrix} \gamma(\omega/c - \beta k_x) \\ \gamma(k_x - \beta\omega/c) \\ k_y \\ k_z \end{pmatrix} \tag{11}$$

where $\beta = v/c$ and $\gamma = 1/\sqrt{1-\beta^2}$, see the derivations in Appendix 3. Correspondingly, the Lorentz transformation is given explicitly by

$$\Lambda = \begin{pmatrix} \gamma & -\beta\gamma & 0 & 0 \\ -\beta\gamma & \gamma & 0 & 0 \\ 0 & 0 & 1 & 0 \\ 0 & 0 & 0 & 1 \end{pmatrix} \tag{12}$$

From the Gauss's law of the Weyl monopole, we apply the Lorentz transformation and eventually we can obtain the following four equations: (see Ref. [38])

$$\begin{aligned} \nabla_{k'} \times \Omega' = 0 &\Rightarrow \begin{cases} \nabla_k \cdot \Upsilon = 0 \\ \nabla_k \times \Omega = \dfrac{\partial \Upsilon}{\partial \omega} \end{cases} \\ \nabla_{k'} \cdot \Omega' = \rho'_W &\Rightarrow \begin{cases} \nabla_k \cdot \Omega = \rho_W \\ \nabla_k \times \Upsilon = -\dfrac{\partial \Omega}{\partial \omega} - j_W \end{cases} \end{aligned} \tag{13}$$

which is the main achievement (1) of our construction of Berry-Maxwell equations in this paper. It is essential to acknowledge that, based on the argument from Feynman's lectures [39], there are additional phenomena beyond the scope of our construction that we cannot explain, even in the classical electrodynamics, such as the radiation reaction effect [40,41]. In addition, we obtain the continuity equation of the reciprocal magnetic-like source and current:

$$\frac{\partial}{\partial \omega} \rho_W + \nabla_k \cdot j_W = 0. \tag{14}$$

We emphasize that the Berry-Maxwell equations do not emerge solely from the geometric phase associated with the Schrodinger equation of matter waves. Instead, we find that the full Berry-Maxwell equations can only be derived from the principles of special relativity and Gauss's law of the Weyl monopoles in parameter space of 4D energy-momentum. The very existence of the monopoles stems from the level crossing between two Floquet-Bloch bands, and is protected due to its topological origin. Consequently, the monopoles cannot be created or destroyed by Lorentz transformations. It is worth noting that while we introduce the Lorentz invariance to the reciprocal Berry-Maxwell field, we do not require the Lorentz invariance for the matter waves. Moreover, it is essential to comprehend that the conventional Maxwell equations can also be derived from the Gauss's law of electric charges and the principles of special relativity [38] (see Appendix 4 for further details).

As an extension of our dual construction, we recognize that Weyl monopoles arise from the level crossing of two nondegenerate Floquet-Bloch bands. Nonetheless, the origin of electric charges in the Maxwell equations is a natural assumption in real spacetime. The fundamental physical origin of electric charge remains elusive even in modern physics. While theorists have explained the electromagnetic forces through gauge invariance when developing the elegant theory of charge quantization by Dirac in 1931 [4], which is related to the coexistence of real-space magnetic monopoles yet to be experimentally observed, it is safe to say that this "quantum origin" does not account for the full Maxwell equations. The reason for this discrepancy lies in our limited understanding of how the Gauss's law of electron charge emerges from matter waves. Instead, our construction offers a clear understanding of the origins of Weyl monopoles in reciprocal electromagnetic field.

In general, we can readily verify that the reciprocal electromagnetic field strength (6) must transform as a second-rank tensor: $\Omega'^{\mu\nu} = \Lambda^{\mu}_{\rho}\Lambda^{\nu}_{\sigma}\Omega^{\rho\sigma}$. Alternatively, in a more concise form, this Berry curvature can be represented as $\Omega' = \Lambda\Omega\Lambda^{\mathrm{T}}$, wherein the indices are hidden. Consequently, by applying the Lorentz transformation, we can effortlessly compute and analyze the generalized Berry connections and curvatures for further implementations. Furthermore, given four quantities $A$ and $Z = (Z_x, Z_y, Z_z)$ and their Lorentz-boosted counterparts $A'$ and $Z' = (Z'_x, Z'_y, Z'_z)$, there exists a relation of the form: $A^2 - \mathbf{Z} \cdot \mathbf{Z} = A'^2 - \mathbf{Z}' \cdot \mathbf{Z}'$. All these quantities have Lorentz invariance transform similar to the spacetime. As a comparison, we list the Lorentz transformation of physical quantities in both energy-momentum space and spacetime, see Table 1 below.

| Quantities in (r,t) | A(r,t) | Z(r,t) | Quantities in (k,ω) | A(k,ω) | Z(k,ω) |
|---|---|---|---|---|---|
| **Position four-vector** | time, ct | position, r | **Four-momentum** | energy (or frequency), E/c | momentum (or wave vector), p |
| **Four-current** | electric charge, ρ | electric current, j | **Reciprocal magnetic four-current** | monopole density, $\rho_W$ | magnetic-like current, $j_W$ |
| **Electromagnetic four-potential** | scalar potential, φ | vector potential, A | **Berry four-connection** | reciprocal scalar potential, χ | reciprocal vector potential, $\mathcal{A}$ |

Table 1: Lorentz transformation of physical quantities in spacetime and dual energy-momentum space, such as position, current, electromagnetic potential and their reciprocal counterparts.

A notable aspect of the construction (Eq. 2) lies in its duality and self-duality when compared to the conventional Maxwell equations, which can be attributed to the concept of Born reciprocity [2]. The self-duality reveals an electric-magnetic duality as follows: $E \rightarrow B, B \rightarrow -E$, and $\Upsilon \rightarrow \Omega, \Omega \rightarrow -\Upsilon$. The duality between Maxwell and Born-Maxwell equations is represented as: $(E, B) \leftrightarrow (\Upsilon, \Omega)$ and the corresponding Born reciprocity $x_\mu \rightarrow p_\mu, p_\mu \rightarrow -x_\mu$. Moreover, we also observe the dual structures of the electromagnetic field and the reciprocal electromagnetic field in their coupling with matter waves ($p_\mu \rightarrow p_\mu - eA_\mu, x_\mu \rightarrow x_\mu - \mathcal{A}_\mu$), see Appendix 5. Despite these dualities, in the following sections we will present three nontrivial observations arising from the Berry-Maxwell equations.

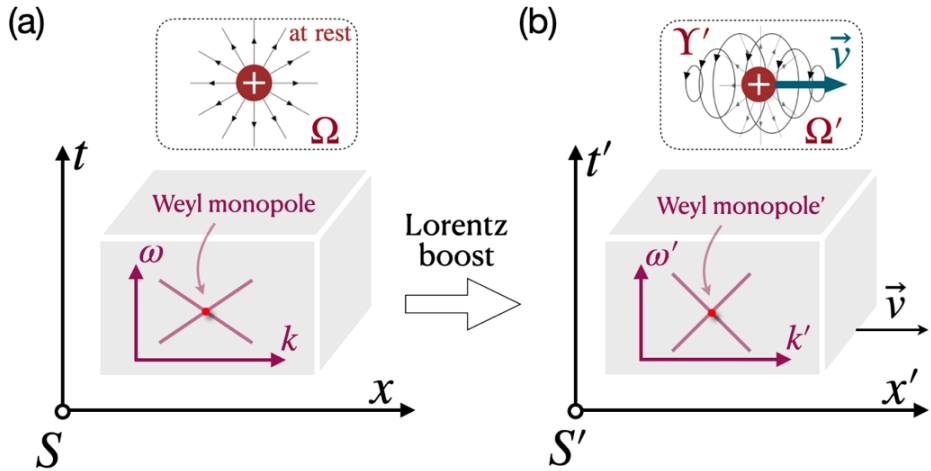

Fig. 2: Lorentz boost of a Weyl monopole in 4D energy momentum space. The Lorentz transformation is applied in the spacetime, and thus applied to the corresponding energy-momentum space. An electric-like Berry curvature ($\Upsilon'$) can be generated from the Lorentz boost, see the inset of (b).

**Lorentz boost of a Weyl monopole.** As depicted in Fig. 2, assuming that the Berry curvature of a Weyl monopole at rest in S frame is given by $\Omega(k) = -\frac{\hat{e}_k}{2k^2}$, the monopole density is given by $\rho_W = \nabla_k \cdot \Omega = 2\pi\delta^3(k - k_0)$, and the electric-like Berry curvature is zero, under the Lorentz transformation, the magnetic- and electric-like Berry curvatures in S' frame become

$$\Omega_{k'_\perp}(k_x = 0) = -\frac{\gamma}{2k'^2_\perp}$$
$$\Upsilon_\varphi(k'_x = 0) = -\frac{\beta}{c}\frac{\gamma}{2k'^2_\perp} \tag{15}$$
$$\Omega_{k_x}(k'_\perp = 0) = -\frac{1}{2\gamma^2 k'^2_x}$$

with the relative velocity $\beta = v/c$, the Lorentz factor $\gamma = 1/\sqrt{1-\beta^2}$, and we set $k_0 = 0$ and $k'^2_\perp = k'^2_y + k'^2_z$. The boost feature of a Weyl monopole is presented in the inset of Fig. 2b.

**Reciprocal Thouless pumping.** It is intriguing to re-examine our reciprocal electromagnetic field in the context of the Chern number and its physical realizations. Figure 3 summarizes a dual correspondence between various physical phenomena and their corresponding Chern numbers across different two-dimensional spaces. Fig. 3a illustrates the Chern number in the x-y plane, exemplified by the superconducting flux quantization [42]. (3b) shows the Chern number in the x-t plane, related to the chiral anomaly [43,44], (3c) presents the Chern number in the $k_x$-$k_y$ plane, pertained to the physics of Quantum Hall effect [19], and (3e) demonstrates the Chern number in the k-t plane (or k-y plane) for Thouless pumping [34]. These mappings underscore the duality between spacetime and energy-momentum space, offering a bird's view of topological effects across various systems.

Specifically, the Chern number in (c) associated with the magnetic-like Berry curvature is given by $C = \frac{1}{2\pi}\oiint \Omega \cdot ds = \frac{1}{2\pi}\oiint \Omega_{k_x,k_y} dk_x dk_y$. Here, the Berry curvature is solely derived from a time-independent system. The aforementioned equation involves an integral over the 2D Brillouin zone of momentum space. This Chern number corresponds to the quantized magnetic flux of superconductors in real space ($C = \frac{1}{2\pi} \oiint B \mathrm{d}x \mathrm{d}y$), and carries significant physical consequences, such as the quantized conductance in QHE [19,20]. They hold dual structure. Building upon our extension, we can also find a Chern number based on the electric-like Berry curvature (see Fig. 3d)

$$C = \frac{1}{2\pi} \oiint \Upsilon(\omega, k) \mathrm{d}\omega \mathrm{d}k. \tag{16}$$

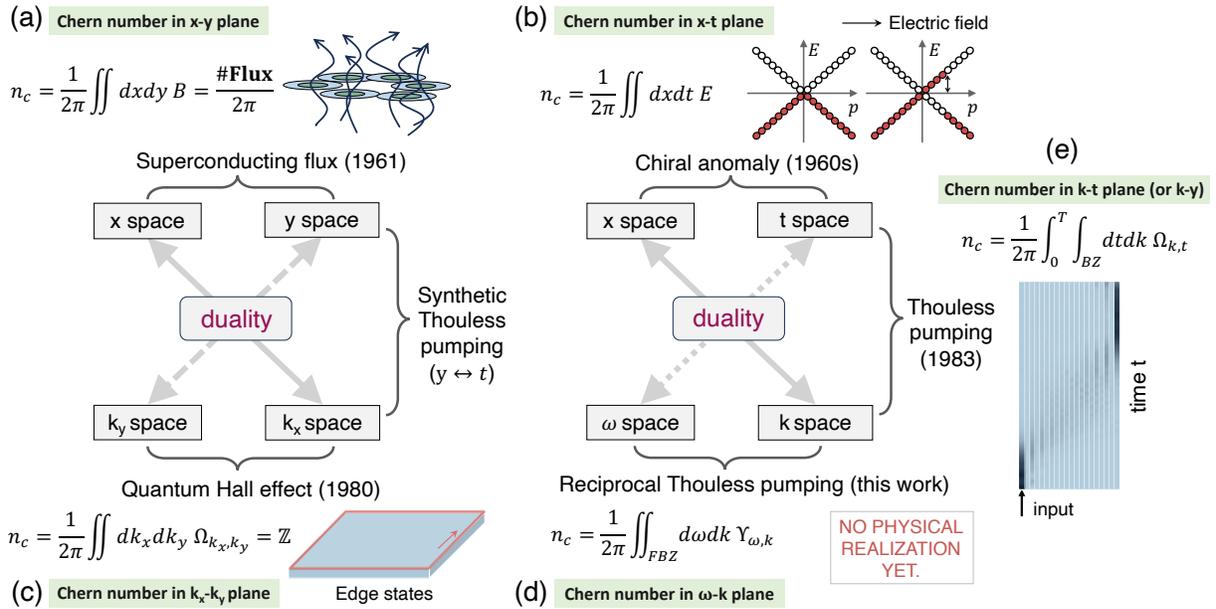

Fig. 3: A gallery of the two-dimensional Chern numbers and their physical realizations. The superconducting magnetic flux (a), chiral anomaly (b), quantum Hall effect (c), Thouless pumping and synthetic Thouless pumping (e) are topological physical effects, which are associated with the Chern numbers in different two-dimensional spaces, respectively. In a dual perspective, our proposed reciprocal Thouless pumping (d) is viewed as a realization of the Chern number in the energy-momentum space.

In this case, the integral is performed over the two-dimensional Floquet-Brillouin zone of energy-momentum space. Although there are several platforms capable of direct measurement of Berry phase and Berry curvature [45–51], a critical question we raise here is whether this electric-like Chern number (or the curvature $\Upsilon$) corresponds to any known or yet unknown physical observation.

Regrettably, to the best of our knowledge, we have not observed any experimental evidence that related to this phenomenon so far, except for a relevant case of the Zak phase only in energy space that has been analyzed in photonic time crystal related to electric-like Berry connection [52]. However, by replacing the frequency ($\omega$) with time ($t$), this Chern number corresponds to Thouless charge pumping [33, 34], described by $C = \frac{1}{2\pi} \oiint \Upsilon(t, k) dt dk$ (Fig. 3e), indicating that one can construct a tight-binding model, such as Rice-Mele model [33], to illustrate the pumping process of the discrete charge per cycle in an adiabatic manner. In addition, from a dual perspective, this electric-like Chern number is a kind of dual quantum anomaly similar to the chiral anomaly [43,44] or the Atiyah-Singer index [53], which has the form $C = \frac{1}{2\pi} \iint E \, dt dx$.

**"Trivial" plane-wave solutions.** The last implementation of our construction is that in the absence of a source or current $\rho_W = 0, j_W = 0$, from (2) we can easily derive $\nabla_k \times (\nabla_k \times \Omega) = -\nabla_k^2 \Omega = \nabla_k \times \left(\frac{\partial \Upsilon}{\partial \omega}\right) = -\frac{\partial^2 \Omega}{\partial \omega^2}$, resulting in the wave equation for the magnetic-like Berry curvature:

$$\frac{\partial^2 \Omega}{\partial \omega^2} - \nabla_k^2 \Omega = 0 \tag{17}$$

Similarly, for the electric-like Berry curvature:

$$\frac{\partial^2 \Upsilon}{\partial \omega^2} - \nabla_k^2 \Upsilon = 0 \tag{18}$$

From these equations, we can obtain the plane-wave solutions $\Omega = \Omega(k, \omega) = e^{-i\tilde{r}k + i\tilde{t}\omega}$, $\Upsilon = \Upsilon(k, \omega) = e^{-i\tilde{r}k + i\tilde{t}\omega}$. Both the electric- and magnetic-like Berry curvatures as "plane waves" in terms of energy and momentum coordinates. A Fourier transformation converts the Berry-Maxwell waves into real spacetime: $\Omega(r, t) = \int d\omega dk^3 \Omega(k, \omega) e^{-ikr - i\omega t} = \delta^3(r - \tilde{r})\delta(t - \tilde{t})$, $\Upsilon(r, t) = \delta^3(r - \tilde{r})\delta(t - \tilde{t})$, where both $\Omega(r, t)$ and $\Upsilon(r, t)$ represent well-defined events in the spacetime at $(\tilde{r}, \tilde{t})$.

**Conclusions.** In this Letter, we present a construction of Berry-Maxwell equations with Lorentz invariance, derived from special relativity and Gauss's law of Weyl monopoles in 4D energy-momentum space. The dual structures of Berry-Maxwell equations are highlighted. The profound implications of the electric-like Berry curvature and the potential for generating reciprocal electromagnetic fields are explored. We propose three significant effects: the Lorentz boost of a Weyl monopole, reciprocal Thouless pumping, and plane-wave solutions. The proposed experimental investigations arising from these frameworks hold the promise of providing invaluable insights into the fundamental properties and physical behaviors of the Berry-Maxwell equations.

**Acknowledgement**

We thank Chenjie Wang, Huaiqiang Wang, Fuchun Zhang, Jian Kang, Daniel Podolsky and Zhaopin Chen for their insightful discussions. We thank the discussions of quantum event wavepacket and its geodynamics with Anzhuoer Li, Shiyue Deng, and also, we appreciate the anomalous referees for clarifying many definitions and statements during the peer-review process. Y.P. is supported by the National Natural Science Foundation of China (Grant No. 2023X0201-417). R.Y. acknowledges the support of Israel Science Foundation's Grant No. 1614/21.

**Appendices:**

**1. Generalization of Berry connections and Berry curvatures in time-dependent Schrödinger equation.**

Suppose that the Hamiltonian varies slowly and parametrically in time, one of the Hamiltonian's eigenstates, $\psi_n$, will evolve into

$$\psi(t) = e^{i\gamma_n(t)} e^{i\theta_n(t)} \psi_n(t) \tag{A1}$$

where the dynamical phase $\theta_n(t) = -\frac{1}{\hbar}\int_0^t E_n(t')dt'$ and the geometric phase $\gamma_n(t) = i\int_0^t \langle \psi_n(t')|\partial_{t'}|\psi_n(t')\rangle dt'$.

Let us say the eigenstate depends on time parametrically (for the relativistic regime, this time is proper time), because there are some system parameters $\mathbf{R}(t) = (\bar{R}_1(t), R_2(t), R_3(t), \cdots)$ in the Hamiltonian which are changing slowly with time

$$\frac{\partial}{\partial t}\psi_n(\mathbf{R}(t)) = \left(\frac{\partial \psi_n}{\partial \bar{R}_1}\right)\frac{\partial \bar{R}_1}{\partial t} + \left(\frac{\partial \psi_n}{\partial R_2}\right)\frac{\partial R_2}{\partial t} + \left(\frac{\partial \psi_n}{\partial R_3}\right)\frac{\partial R_3}{\partial t} + \cdots = (\partial_\mu \psi_n)\frac{\partial R_\mu}{\partial t} \tag{A2}$$

It is noteworthy that we define the parameter space $\mathbf{R}(t)$, which possesses a symplectic structure in order to engage with the constraints from the special relativity. Specifically, we have the flexibility to choose the parameter space $\mathbf{R}(t)$ as either the energy-momentum space $(\omega, k_x, k_y, k_z)$ or the spacetime $(t, x, y, z)$ [Note that the original time in the temporal evolution of the system corresponds to the proper time, sometimes, denote as $\tau$, and thus, $H(\tau) = H(\mathbf{R}(\tau))$]. Consequently, we can identify special parameter spaces that exhibit Lorentz invariance under the Lorentz transformation. So that the geometric phase

$$\gamma_n(T) = i\oint \langle \psi_n|\partial_{R_\mu}\psi_n\rangle dR_\mu \tag{A3}$$

Note that this Berry phase is different from the conventional geometric phase [24] which does not require the Lorentz invariance on the parameter space.

Now, we can define the Berry connection and Berry curvature in analogy with the four-potential and electromagnetic field tensor:

$$\mathcal{A}_\mu = i \langle \psi_n | \partial_{R_\mu} \psi_n \rangle$$
$$\Omega_{\mu\nu} = \partial_\nu \mathcal{A}_\mu - \partial_\mu \mathcal{A}_\nu \quad (A4)$$

so that the Berry phase (A3)

$$\gamma_n(T) = \oint_\mathcal{C} \mathcal{A}_\mu \, dR_\mu = \iint_\mathcal{S} \Omega_{\mu\nu} \, dR_\mu dR_\nu \quad (A5)$$

where the surface $\mathcal{S}$ is the surface circulated by the loop $\mathcal{C}$. Later, we would see that the measure $dR_\mu dR_\nu$ is an invariance under Lorentz transformation.

**2. Construction of the Gauss's law of Weyl points as the source of the magnetic-like Berry curvature.**

In order to construct a magnetic monopole in momentum space as a source for generating Berry-Maxwell equations, for a time-independent quantum system we can express the Berry curvature $\Omega_{\mu\nu}$ in terms of

$$\Omega_{\mu\nu} = i \langle \partial_{R_\mu} \psi_n | \partial_{R_\nu} \psi_n \rangle - i \langle \partial_{R_\nu} \psi_n | \partial_{R_\mu} \psi_n \rangle \quad (B1)$$

as we insert that the completeness $\sum_{n'} |\psi_{n'}\rangle\langle\psi_{n'}| = I$, and then we obtain

$$\Omega_{\mu\nu} = i \sum_{n'} \langle \partial_{R_\mu} \psi_n | \psi_{n'} \rangle \langle \psi_{n'} | \partial_{R_\nu} \psi_n \rangle - (\mu \leftrightarrow \nu)$$
$$= i \sum_{n' \neq n} \frac{\langle n | \partial_{R_\mu} H | n' \rangle \langle n' | \partial_{R_\nu} H | n \rangle}{(E_n - E_{n'})^2} - (\mu \leftrightarrow \nu) \quad (B2)$$

Now, we notice that the term $n' = n$ is cancelled out by the symmetry under $(\mu \leftrightarrow \nu)$, where $n, n'$ represent the abbreviated notation for eigenstates $\psi_n, \psi_{n'}$, respectively.

We notice that the Berry curvature will diverge at the degenerate point of two bands where $E_n = E_{n'}$. In order to see the power law of the divergence, we linearize the dominator at the degenerate point when $R = R_c$:

$$E_n - E_{n'} = \left(E_n(R_c) + \frac{\partial E_n}{\partial R}(R - R_c)\right) - \left(E_{n'}(R_c) + \frac{\partial E_{n'}}{\partial R}(R - R_c)\right) \quad (B3)$$
$$= \left(\frac{\partial E_n}{\partial R} - \frac{\partial E_{n'}}{\partial R}\right)(R - R_c)$$

Under the conditions that $\left(\frac{\partial E_n}{\partial R} - \frac{\partial E_{n'}}{\partial R}\right) \neq 0$ and the numerator are nonzero, we can expect:

$$\Omega \sim \frac{1}{(R - R_c)^2} \quad (B4)$$

Subsequently, a nontrivial Gauss's law with a power law behavior emerges: $\nabla_R \cdot \Omega \sim \delta^3(k - k_c)$. It is important to highlight that at the degenerate point where $E_n = E_{n'}$, the adiabatic approximation fails, leading to that the Berry connection is singularity ill-defined. However, the Berry curvature remains comprehensible by taking the divergence as a magnetic monopole in the parameter space.

To be specific, let us take an example (see Ref. [37]). A 3D Weyl point has a Hamiltonian $H_{eff} = v_F k \cdot \sigma$ in momentum space. One of the eigenstates is given by

$$|u_+(k, \theta, \phi)\rangle = \begin{pmatrix} \sin\frac{\theta}{2} \\ -\cos\frac{\theta}{2} e^{i\phi} \end{pmatrix} \quad (B5)$$

with $\cos\theta = \frac{k_z}{k}$, $k = \sqrt{k_x^2 + k_y^2 + k_z^2}$. The Berry connection is $\mathcal{A} = -i\langle u_n(k,\omega,x,t)|\nabla_k|u_n(k,\omega,x,t)\rangle$ and the derivative is $\nabla_k = \left(\partial_k, \frac{1}{k}\partial_\theta, \frac{1}{k\sin\theta}\partial_\phi\right)$, so that

$$(\mathcal{A}_k, \mathcal{A}_\theta, \mathcal{A}_\phi) = \left(0, 0, \frac{\cos^2\frac{\theta}{2}}{k\sin\theta}\right) \quad (B6)$$

and correspondingly, the Berry curvature is $\Omega = \nabla_k \times \mathcal{A} = -\frac{\hat{e}_k}{2k^2}$.

## 3. Lorentz invariance of the interval between spacetime and the phase of a plane wave

In general, we address that the choice of parameter space is arbitrary. But if the parameter space is spacetime or energy-momentum space, we would be glad to introduce the Lorentz constraints on the parameter spaces. The two postulates of the special relativity naturally give rise to the Lorentz transformation, which can be interpreted as a boost transformation of spacetime from S frame to S' frame (see Fig. 1)

$$\begin{pmatrix} ct' \\ x' \\ y' \\ z' \end{pmatrix} = \begin{pmatrix} \gamma & -\beta\gamma & 0 & 0 \\ -\beta\gamma & \gamma & 0 & 0 \\ 0 & 0 & 1 & 0 \\ 0 & 0 & 0 & 1 \end{pmatrix} \begin{pmatrix} ct \\ x \\ y \\ z \end{pmatrix} = \begin{pmatrix} \gamma ct - \beta\gamma x \\ \gamma x - \beta\gamma ct \\ y \\ z \end{pmatrix} \quad (C1)$$

with $\beta = v/c$ and $\gamma = 1/\sqrt{1-\beta^2}$. Correspondingly, the inverse Lorentz transformation

$$\begin{cases} t = \gamma\left(t' + \frac{vx'}{c^2}\right) \\ x = \gamma(x' + vt') \\ y = y' \\ z = z' \end{cases} \quad (C2)$$

We know that the interval $ds^2 = c^2 dt^2 - dx^2 - dy^2 - dz^2 = c^2 dt'^2 - dx'^2 - dy'^2 - dz'^2$ is invariant under the Lorentz transformation. So this interval is a Lorentz invariance. We can also construct a gap invariance $dm^2 = d\omega^2 - c^2 dk^2$, which implies the energy-momentum transform under Lorentz transformation similar to the transformation of spacetime coordinates:

$$\begin{pmatrix} \omega'/c \\ k'_x \\ k'_y \\ k'_z \end{pmatrix} = \begin{pmatrix} \gamma & -\beta\gamma & 0 & 0 \\ -\beta\gamma & \gamma & 0 & 0 \\ 0 & 0 & 1 & 0 \\ 0 & 0 & 0 & 1 \end{pmatrix} \begin{pmatrix} \omega/c \\ k_x \\ k_y \\ k_z \end{pmatrix} = \begin{pmatrix} \gamma(\omega/c - \beta k_x) \\ \gamma(k_x - \beta\omega/c) \\ k_y \\ k_z \end{pmatrix} \quad (C3)$$

Now, let us check a phase term:

$$k' \cdot r' - \omega' t' = \gamma\left(k_x - \frac{\beta\omega}{c}\right)(\gamma x - \beta\gamma ct) - \cdots = k \cdot r - \omega t = \text{invariants} \quad (C4)$$

So, the phase of a plane wave is also a Lorentz invariant, which indicates the consistence between Fourier transformation and Lorentz transformation, even for the Maxwell equations in coupling with the nonrelativistic Schrödinger equation. The Lorentz transformation (C3) between frequency (energy) and wavevector (momentum) can be experimentally confirmed by the Doppler effect and the aberration effect.

**4. Constructing Maxwell equations by introducing Lorentz invariance and Gauss's law of electric charge.**

Even though we have embraced the validity of the Maxwell equations, it remains pertinent to explore the derivation, as well as the limitations, of these equations through the gauge invariance of the Schrödinger equation. To this end, we proceed by constructing the 4-form vector potential in 3+1D spacetime. We know that the dynamics of a quantum particle moving in the presence of electromagnetic field, is governed by

$$i\hbar \frac{\partial}{\partial t}\psi = H\psi \qquad (D1)$$

with the minimal coupling $H = \frac{1}{2m}\left(\frac{\hbar}{i}\nabla - eA\right)^2 + e\phi$. And there is a gauge degree of freedom:

$$\begin{cases} A' = A + \nabla\Lambda \\ \phi' = \phi - \frac{\partial}{\partial t}\Lambda \end{cases} \qquad (D2)$$

and meanwhile the wavefunction satisfies

$$\psi' = e^{\frac{ie\Lambda}{\hbar}}\psi, \qquad (D3)$$

We find that the new equation has the same structure:

$$\begin{aligned} i\hbar \frac{\partial}{\partial t}\psi &= \left(\frac{1}{2m}\left(\frac{\hbar}{i}\nabla - eA\right)^2 + e\phi\right)\psi \\ i\hbar \frac{\partial}{\partial t}\psi' &= \left(\frac{1}{2m}\left(\frac{\hbar}{i}\nabla - eA'\right)^2 + e\phi'\right)\psi' \end{aligned} \qquad (D4)$$

This gauge invariance is important in quantum mechanics, and it offers an alternative way to define the vector potential. Let us assume: $A' = A + \nabla \Lambda = 0, \phi' = \phi - \frac{\partial}{\partial t}\Lambda = 0$, so that we have

$$A = -\nabla \Lambda, \phi = \frac{\partial}{\partial t}\Lambda \tag{D5}$$

Hence, we obtain $\Lambda = \Lambda(r, t)$. From the second equation in (D4) with $A' = 0, \phi' = 0$, the new wavefunction $\psi'$ has a plane wave solution: $\psi' = \psi_0 e^{\frac{i(pr - E_p t)}{\hbar}}$ with the energy dispersion $E_p = \frac{p^2}{2m}$. By employing this plane-wave solution, we subsequently solve the original wavefunction

$$\psi = e^{-\frac{ie\Lambda(r,t)}{\hbar}}\psi' = e^{-\frac{ie\Lambda(r,t)}{\hbar}} e^{\frac{i(pr-E_p t)}{\hbar}}\psi_0 = e^{\frac{i(pr-E_p t)}{\hbar}} u(r,t) \tag{D6}$$

In this sense, we can rewrite (D5) as:

$$\begin{aligned} \phi &= \frac{\partial}{\partial t}\Lambda = \left(\frac{i\hbar}{q}\right)\langle u(r,t)|\partial_t|u(r,t)\rangle \\ A &= -\nabla\Lambda = -\left(\frac{i\hbar}{q}\right)\langle u(r,t)|\nabla|u(r,t)\rangle \end{aligned} \tag{D7}$$

Furthermore, we redefine the electric and magnetic fields

$$\begin{cases} E = -\frac{\partial}{\partial t}\vec{A} - \nabla\phi = \frac{\partial}{\partial t}(\nabla\Lambda) - \nabla\left(\frac{\partial}{\partial t}\Lambda\right) = 0 \\ B = \nabla \times \vec{A} = -\nabla \times (\nabla\Lambda) = 0 \end{cases} \tag{D8}$$

The occurrence of the first zero arises due to the interchange of the spatial and time derivatives, while the second zero emerges from the property that the curl of a gradient is zero. This is reasonable, as it highlights the fact that, the mere application of a local gauge transformation from the time-dependent Schrödinger equation (TDSE), is insufficient to derive the complete set of conventional Maxwell equations.

Indeed, we can derive the Maxwell equations by directly employing the principles of special relativity and Gauss's law of electric charge. Specifically, when considering an electric charge in the moving frame S', it satisfies the electrostatic equation:

$$E' = \frac{1}{4\pi\epsilon_0} \frac{\hat{e}_{r'}}{r'^2} \tag{D9}$$

so that we can easily obtain

$$\begin{cases} \nabla \times E' = 0 \\ \nabla \cdot E' = \dfrac{\rho'}{\epsilon_0} \end{cases} \tag{D10}$$

By applying the Lorentz transformation to (D10) into the rest frame S, Haskell's et al. [38] has shown that these two equations transform into the four equations:

$$\nabla \times E' = 0 \Rightarrow \begin{cases} \nabla \cdot B = 0 \\ \nabla \times E = -\dfrac{\partial}{\partial t} B \end{cases}$$

$$\nabla \cdot E' = \dfrac{\rho'}{\epsilon_0} \Rightarrow \begin{cases} \nabla \cdot E = \dfrac{\rho}{\epsilon_0} \\ \nabla \times B = \mu_0 J + \epsilon_0 \mu_0 \dfrac{\partial}{\partial t} E \end{cases} \tag{D11}$$

The Maxwell equations can be expressed as:

$$\begin{cases} \dfrac{\partial F_{\mu\nu}}{\partial x_\lambda} + \dfrac{\partial F_{\nu\lambda}}{\partial x_\mu} + \dfrac{\partial F_{\lambda\mu}}{\partial x_\nu} = 0 \\ \dfrac{\partial F_{\mu\nu}}{\partial x_\nu} = \mu_0 J_\mu \end{cases} \tag{D12}$$

where $F_{\mu\nu}$ is the electromagnetic field tensor, and $J_\mu = (\rho, J)$ is the four-current following the continuity equation $\partial_\mu J_\mu = 0$. In general, the conventional EM field tensor is explicitly given by

$$F_{\mu\nu} = \partial_\nu \mathcal{A}_\mu - \partial_\mu \mathcal{A}_\nu = \begin{pmatrix} 0 & +E_x & +E_y & +E_z \\ -E_x & 0 & -B_z & +B_y \\ -E_y & +B_z & 0 & -B_x \\ -E_z & -B_y & +B_x & 0 \end{pmatrix} \tag{D13}$$

Here, we demonstrate why the Maxwell equations cannot be derived from the gauge transformation for TDSE. We then derive the conventional Maxwell equations in four-dimensional spacetime from Lorentz invariance and Gauss's law of electric charge. This construction reveals a dual structure with our formulation of the Berry-Maxwell equations.

## 5. The phase-space wavepacket in presence of both Maxwell and Berry-Maxwell equations.

In this section, we compare the on-shell and off-shell Floquet-Bloch states. For our setup, the on-shell and off-shell states are analytically similar. We introduce a "proper time" to relax the energy-momentum dispersion constraint [29, 54]. Starting from the time-dependent Schrodinger equation (TDSE) in standard form as:

$$\mathcal{L}(-i\partial_x, i\partial_t; x, t)\psi = 0, \quad (E1)$$

where the operator $\mathcal{L}(\partial_t, \partial_x; t, x) = H - i\hbar\partial_t = H_F$ is the Floquet Hamiltonian. By applying the Floquet-Bloch theorem, $\psi(x,t) = e^{-i\omega t + ikx}u(x,t)$, we obtain the dispersion relation by solving the following equation: $\mathcal{L}(-i\partial_x + k, i\partial_t - \omega; x, t)u(x,t) = 0$. Thus, we have the dispersion $\omega = E(k)$. To relax the constraint, we extend the TDSE (E1) to the geodynamic equation:

$$\mathcal{L}(-i\partial_x, i\partial_t; x, t)\Psi(x, t) = i\hbar \frac{\partial}{\partial \tau}\Psi(x, t) \quad (E2)$$

where $\tau$ is an introduced "proper time" representing quantum state evolution. Substituting $\Psi(x,t) = e^{-\frac{i\lambda_0 \tau}{\hbar}}\psi(x,t)$ yields the eigenproblem:

$$\mathcal{L}(-i\partial_x, i\partial_t; x, t)\psi = \lambda_0 \psi. \quad (E3)$$

The original equation corresponds to the zero "energy" solution: $\lambda_0 = 0$. Applying the Floquet-Bloch theorem again gives a dispersion function $\lambda_0 = \lambda_0(k, \omega)$ [29]. The off-shell Floquet-Bloch states $u_{\lambda_0}(x,t)$ satisfy $\mathcal{L}(-i\partial_x + k, i\partial_t - \omega; x, t)u_{\lambda_0}(x,t) = \lambda_0 u_{\lambda_0}(x,t)$, with the on-shell constraint $\omega = \omega(k)$ arising when $\lambda_0 = 0$. It indicates the physical world lies in the shell $\lambda_0 = 0$. Rewriting the geodynamic equation (E2) as:

$$(\mathcal{L}(-i\partial_x, i\partial_t; x, t) - i\hbar\partial_\tau)\Psi(x,t) = (H(\partial_x, x, t) - i\hbar\partial_t - i\hbar\partial_\tau)\Psi(x,t) = 0 \quad (E4)$$

Here, the geodynamic equation has two time scales: physical time for periodic modulation and the introduced "proper time". The proper time $\tau$ corresponds to the spatiotemporal deformation of spacetime crystals, which is analogous to these additional 'slow' time variables in solving ODE and PDE systems using multiple scales [55]. For our purposes, the Floquet Hamiltonian is independent of proper time, leading to:

$$\mathcal{L}(-i\partial_x + k, i\partial_t - \omega - \lambda_0; x, t)u_{\lambda_0}(x,t) = 0 \quad (E5)$$

The off-shell Floquet-Bloch state $u_{\lambda_0}(x,t)$ is analytical similar to the on-shell Floquet-Bloch state $u(x,t)$ with a frequency shift of $\lambda_0$. Following Qian Niu et al. [28, 37], we define a quantum event wavepacket in eigen-dimension phase space:

$$|w(k_0, \omega_0, x, t)\rangle = \int dk^3 d\omega \, a(k, \omega)|u(k, \omega, x, t)\rangle \quad (E6)$$

where $a(k, \omega)$ is a localized function in energy-momentum space centered at $(\omega_0, k_0)$. Analogous to classical mechanics, $a(k, \omega)$ represents a delta function representing an event in special relativity. We couple both the Maxwell electromagnetic field and the reciprocal Maxwell electromagnetic field through the generalized minimal couplings:

$$\begin{cases} \hbar k \to \hbar k - eA(x,t) \\ \hbar \omega \to \hbar \omega - e\phi(x,t) \\ x \to x - \mathcal{A}(k,\omega) \\ t \to t - \chi(k,\omega) \end{cases} \quad (E7)$$

The semiclassical motion of equation of the wavepacket in phase space along the worldline with a proper time $\tau$:

$$\begin{cases} \dot{t} = -\dfrac{\partial \lambda_0}{\partial \omega} - \Upsilon \cdot \dot{k} \\ \dot{x} = \dfrac{\partial \lambda_0}{\partial k} - \Omega \times \dot{k} - \Upsilon \dot{\omega} \\ \dot{\omega} = \dfrac{\partial \lambda_0}{\partial \tau} - eE \cdot \dot{x} \\ \dot{k} = -\dfrac{\partial \lambda_0}{\partial x} - B \times \dot{x} - eE\dot{t} \end{cases} \quad (E8)$$

It is worth noting that equation (E6) is not merely an extension of the semiclassical phase space wavepacket from reference [35]. As we explicitly pointed out, the Berry-Maxwell equations can only be obtained by introducing Einstein's Lorentz covariance into the momentum-energy space, rather than solely induced from the governing wave equation. Therefore, Eq. E6 represents the dynamics of a quantum event wavepacket in the presence of both the electromagnetic field $(E, B)$ and the reciprocal electromagnetic field $(\Upsilon, \Omega)$, which satisfy the Maxwell equations and the Berry-Maxwell equations, respectively.